# An Interpretable Mapping from a Communication System to a Neural Network for Optimal Transceiver-Joint Equalization

Zhiqun Zhai, Hexun Jiang, Mengfan Fu, Lei Liu, Lilin Yi, Weisheng Hu, and Qunbi Zhuge*

*Abstract*— In this paper, we propose a scheme that utilizes the optimization ability of artificial intelligence (AI) for optimal transceiver-joint equalization in compensating for the optical filtering impairments caused by wavelength selective switches (WSS). In contrast to adding or replacing a certain module of existing digital signal processing (DSP), we exploit the similarity between a communication system and a neural network (NN). By mapping a communication system to an NN, in which the equalization modules correspond to the convolutional layers and other modules can been regarded as static layers, the optimal transceiver-joint equalization coefficients can be obtained. In particular, the DSP structure of the communication system is not changed. Extensive numerical simulations are performed to validate the performance of the proposed method. For a 65 GBaud 16QAM signal, it can achieve a 0.76 dB gain when the number of WSSs is 16 with a -6 dB bandwidth of 73 GHz.

## I. INTRODUCTION

WITH the rapid progress of 5G, the internet of things, cloud computing, and high definition video, the demand for the capacity of optical networks is exponentially increasing. In this context, elastic optical network (EON) is expected to increase the capacity and flexibility of optical networks. In EON, flex-grid wavelength selective switch (WSS) which performs channel routing on a wavelength basis is a key device. Since optical communication systems are evolving towards higher baud rate and higher spectral efficiency, the guard band between adjacent channels is decreasing. As a consequence, the optical filtering caused by cascaded WSSs, leading to inter-symbol interference (ISI), has become a nonnegligible factor that influences the capacity of transmission.

The ISI can be compensated using a digital post-compensation filter in a coherent receiver (Rx) [1-2]. However, the digital filter at the Rx side enhances noise power since low-power frequency components are emphasized during the ISI compensation, resulting in signal-to-noise ratio (SNR) degradation. It has been demonstrated that the degradation can be alleviated by transceiver-joint equalization [3-5], and various schemes have been proposed for mitigating the filtering penalties. In [3], the Rx-side adaptive equalizers directly serve as a time-domain pre-equalization scheme. However, it is not optimal from a theoretical perspective. In [4], the pre-equalization filter coefficients are theoretically derived. They assume the optimal equalization can be achieved when all inter-symbol interference is compensated. However, this is not valid since noise always exists in a realistic system. Furthermore, they do not consider the excess power loss, which is induced by larger power allocation to lossy frequency components. Other compensation techniques such as partial pre-emphasis have been studied [5] which are based on empirical or brute-force methods. However, it only searches in a one-dimensional space which cannot be proved to be optimal and it is impractical to search in all dimensions with limited calculation resources. Meanwhile, since the link configurations in a network are quite diverse, it is too complex to optimize every link based on brute-force methods.

Analytical model-based optimization may be a solution to achieve the optimal transceiver-joint equalization, in which objective function and constraints of the optimization are required. First, the constraints should be established to describe the entire propagation model including the transmitter (Tx), transmission link and Rx. Second, the objective function is the performance of the system and the target of the optimization is to achieve the best performance in certain metrics such as minimum mean square error (MMSE). In the scenario of transceiver-joint equalization, since the signal suffers from cascaded impairments such as fiber nonlinearity, optical filtering and noise, the constraints are complicated, making it very tough to use a traditional optimization tool to solve this

This paper is submitted for review on February 9, 2021. This work was supported in part by the National Key R&D Program of China under Grant 2018YFB1800902, in part by the National Natural Science Foundation of China under Grant 61801291, in part by the Shanghai Rising-Star Program under Grant 19QA1404600 *(Corresponding author: Qunbi Zhuge.)*

The authors are with the State Key Laboratory of Advanced Optical Communication Systems and Networks, Shanghai Institute for Advanced Communication and Data Science, Shanghai Jiao Tong University, Shanghai 200240, People's Republic of China. (e-mail: zhaizhiqun@sjtu.edu.cn; jianghexun@sjtu.edu.cn; mengfan.fu@sjtu.edu.cn; liulei_sjtu@sjtu.edu.cn; lilinyi@sjtu.edu.cn; wsh@sjtu.edu.cn; qunbi.zhuge@sjtu.edu.cn ).

problem.

In recent years, artificial intelligence (AI) techniques such as neural networks (NN) have made rapid progress for their abilities to approximate nonlinear functions and perform optimization. The applications of AI in communication systems are mainly focused on using NN for nonlinear equalization or optimal constellation design [6-14], in which an NN is used as an individual digital signal processing (DSP) block (e.g. equalization or decoding). However, since an NN is a kind of 'black box', its interpretability is relatively poor, limiting its adoption in practical communication systems.

In this work, we introduce an interpretable mapping from an optical communication system to an NN, which paves the way to leverage the optimization ability of AI without introducing a "black box". In this scheme, the pre-equalization filter at the Tx and the adaptive filter at the Rx correspond to two one-dimension convolutional layers whose weights need to be updated while the other DSP blocks can be regarded as static layers. By optimizing the NN, the optimal transceiver-joint equalization coefficients can be obtained. Since the equalization coefficients of Rx can be obtained adaptively in a practical system, only the coefficients of the pre-equalization need to be extracted while the DSP structure of the communication system is not changed. Extensive simulations are conducted to verify the proposed scheme, and up to 0.76 dB SNR improvement can be obtained for 65 GBaud 16-quadrature amplitude modulation (16-QAM) signals. The link contains 16 cascaded WSSs with a 6 dB bandwidth of 73 GHz, and various fiber impairments including nonlinearity, chromatic dispersion (CD) and polarization effects are considered.

The rest of the paper is organized as follows. Section II introduces the mapping from an optical communication to an NN. Afterwards, the process of obtaining the optimal transceiver-joint equalization coefficients along with the training procedure is described. In Section III, we present the simulation setup in detail, and discuss the performance of the proposed scheme in different scenarios. The paper is concluded in Section IV.

## II. PRINCIPLE

### A. Optical Communication Model and the Mapped NN

As shown in Fig. 1(a), a typical optical communication system consists of coding, Tx DSP, channel, Rx DSP and decoding modules. In order to improve the performance of the system, every parameter of these modules need to be optimized. However, a practical communication system is generally complex, making it difficult to optimize, especially when some parameters need to be jointly optimized. Since AI techniques have powerful optimization abilities, they provide a promising way to overcome this difficulty. However, the poor interpretability prevents their adoptions in practical systems.

We propose to map the communication system to an NN, which paves the way to leverage the optimization ability without introducing a black box. Since the optical filtering impairment is linear, we only consider the mapping of the linear equalization module which is a finite impulse response (FIR) filter while other modules are considered as the static layers of an NN as illustrated in Fig. 1(b). In the future, more modules can be mapped to certain layers of an NN as shown in Fig. 1(c) by exploring the similarity between communication systems and NN, and then more parameters can be jointly optimized to achieve an even better performance. In the following, an interpretable mapping from a FIR filter to a convolutional layer is introduced.

In this section, the principle is described with only one polarization for simplicity, but the simulations in the next section is conducted with dual-polarization signals. We denote the input signal and the coefficients of the FIR filter as $s(n)$ and $h(n)$, respectively. The output of the FIR filter $r(n)$ can be expressed as

$$r(n) = s(n)*h(n) = \sum_{i=-\infty}^{\infty} s(i) \cdot h(n-i) \quad (1)$$

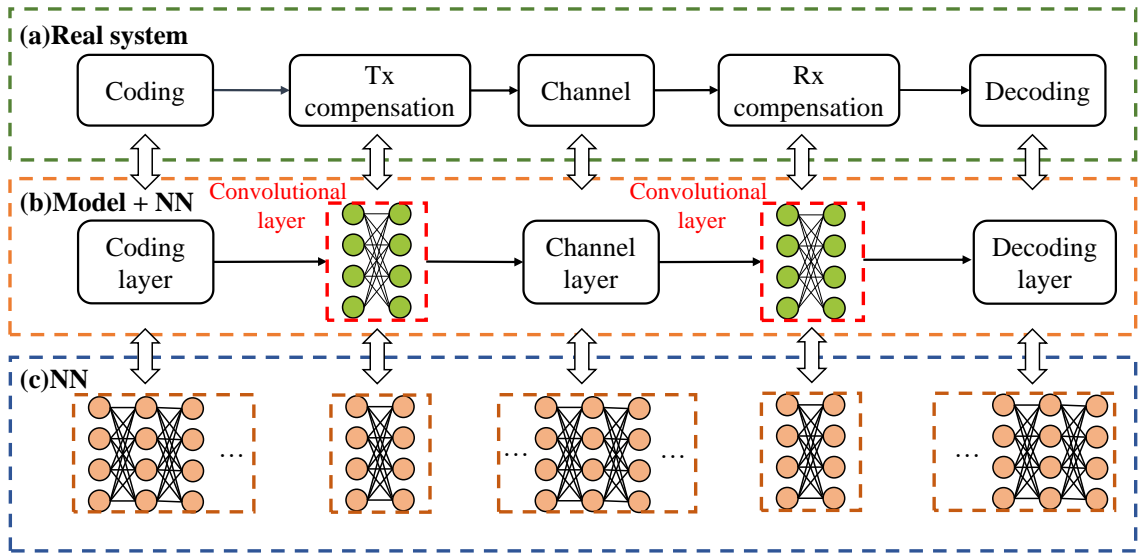

Fig. 1. The schematic of the scheme

where $*$ denotes convolution in the conventional DSP. The function of a one-dimension convolutional layer is defined as

$$out(N_i, C_{out_j}) = bias(C_{out_j}) + \sum_{k=0}^{C_{in}-1} weight(C_{out_j}, k) \odot input(N_i, k) \quad (2)$$

where the dimensions of the input and the output are $(N, C_{in}, L)$ and $(N, C_{out}, L_{out})$, respectively. $N$ is a batch size, $C$ is the number of channels, and $L$ is the length of signal sequence. $\odot$ represents a cross-correlation operation. When the number of input channels is 1 and the bias is 0, the output of the convolutional layer can be expressed as

$$out = weight \odot input \quad (3)$$

According to the definition, the cross-correlation can be expressed as

$$r_{sh} = s(n) \odot h(n) = \sum_{n=-\infty}^{\infty} s(n)h(n-\tau) = \sum_{i=-\infty}^{\infty} s(i)h(-(\tau-i))$$
$$= s(n) * h(-n) \quad (4)$$

From (4), it is proved that a FIR filter can be represented by a convolutional layer and the weights of the convolutional layer flip the coefficients of a FIR filter. The parameters of the convolutional layer are defined in Table I.

For a FIR filter with complex coefficients, the input is denoted as

$$x(n) = x_r(n) + jx_i(n) \quad (5)$$

And its coefficients are denoted as

$$h(n) = h_r(n) + jh_i(n) \quad (6)$$

TABLE I
THE DEFINITION OF THE PARAMETERS OF A CONVOLUTION LAYER

| Parameter | Definition |
| --- | --- |
| Input channels | Number of channels in the input |
| Output channels | Number of channels produced by the convolution |
| Kernel size | Size of the convolving kernel |
| Padding | Zero-padding added to both sides of the input |
| Groups | Number of blocked connections from input to output channels |

Therefore, the output of the FIR filter with complex coefficients can be obtained as

$$y(n) = x(n) * h(n) = y_r(n) + jy_i(n) \quad (7)$$

where $y_r(n)$ and $y_i(n)$ are the real and imaginary parts of the output signal, respectively, holding the relationship as

$$y_r(n) = x_r(n) * h_r(n) - x_i(n) * h_i(n)$$
$$y_i(n) = x_r(n) * h_i(n) + x_i(n) * h_r(n) \quad (8)$$

As NNs are designed primarily for real valued data, the complex input sequence is treated as two real-valued sequences. Therefore, we set the 'input channels' to 2, one is for the real parts of the signal and the other for the imaginary parts of the signal. The 'output channels' is set to 1. The weights and the input of the convolutional layer are denoted as $[h_1(n), h_2(n)]$ and $[x_1(n), x_2(n)]$. From (2) we can obtain the output of the convolutional layer which can be expressed as

$$y = x_1(n) \odot h_1(n) + x_2(n) \odot h_2(n)$$
$$= x_1(n) * h_1(-n) + x_2(n) * h_2(-n) \quad (9)$$

Comparing (9) with (8), as expected the FIR filter with complex coefficients can also be represented by a convolutional layer. When the input of the convolutional layer is $[x_r(n); -x_i(n)]$, the output is $y_r$, and when the input is $[x_i(n); x_r(n)]$, the output is $y_i$. The corresponding parameters are summarized in Table II.

TABLE II
PARAMETERS OF A FIR FILTER AND A CONVOLUTIONAL LAYER

| Parameters | FIR filter | One-dimension convolutional layer |
| --- | --- | --- |
| Input | $x_r(n)+jx_i(n)$ | $[x_r(n); -x_i(n)]$ |
|  |  | $[x_i(n); x_r(n)]$ |
| Output | $y_r(n)+jy_i(n)$ | $[y_r(n); y_i(n)]$ |
| Filter weight | $[h_r(n)+jh_i(n)]$ | $[h_r(-n); h_i(-n)]$ |
| Filter length | Filter_length | Kernel_size |
| Padding | ------ | Kernel_size/2 |
| Groups | ------ | 1 |

*B. The framework of the NN-based optimization*

In the aforementioned analysis, we have introduced an interpretable mapping from a communication system to an NN for transceiver-joint equalization, in which the FIR filters including the Tx pre-equalization filter and Rx adaptive filter are equivalent to the convolutional layers of the NN. In this section, the process of the NN-based optimization is presented. The schematic is shown in Fig. 2.

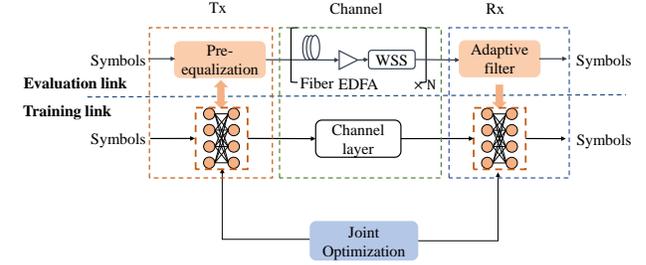

Fig. 2. The schematic of the NN-based transceiver-joint equalization.

The first step is to build an NN based on the mapping. The Tx pre-equalization filter and Rx adaptive filter are replaced with two convolutional layers whose weights need to be updated while channel, transceiver and other DSP can be regarded as the static layers of the NN, and the functions are unchanged and the same as those in traditional communication system. The inputs of the NN are the symbols before pre-equalization and the outputs of NN are the symbols after post-equalization. Since the performance and the convergence speed are related to the parameter initialization, choosing a suitable initial set of parameters is important to facilitate successful training. We set the initial parameters of the first convolutional layer to a matrix whose middle coefficient is 1 and others are small values close to 0. The initial parameters of the last convolutional layer are set to the converged coefficients of the adaptive filter without pre-equalization. In this case, the initial performance of the NN is close to the performance of the communication system without pre-equalization.

After building the NN, the second step is to obtain the optimal transceiver-joint equalization filter coefficients by training the NN. The goal of training the NN is to minimize the mean square error (MSE) between the output symbol $\hat{x}$ and the input symbol $x$ which is utilized as the loss function. The loss function can be expressed as

$$L = ||\hat{x} - x||^2 \qquad (10)$$

The training process consists of iteratively varying a set of trainable parameters, which is updated by a gradient descent algorithm, and evaluating the performance of the system in terms of MSE loss. The weights of the first and second convolutional layers are denoted as $w^0$ and $h^0$, respectively. The Adam optimization, which is a stochastic gradient descent method based on adaptive estimation of the first-order and second-order moments, is used to update the trainable parameters. The details of the training procedure are described in Algorithm 1.

---
**Algorithm 1** End-to-end training procedure
---
1. Generate a training dataset and a testing dataset
2. Initialize the trainable parameters of the two convolutional layers: $w^0$ and $h^0$
3. For n → 1, …, $N_{iter}$ do
    1) Generate random symbols {x} from the dataset
    2) Upsample and pulse shape the symbols to get $\{x(\tau)\}$
    3) Pre-equalize the signal using the first convolutional layer and get $\{s(\tau)\}$
    4) Propagate $\{s(\tau)\}$ in the channel to get $\{r(\tau)\}$
    5) Detect $\{r(\tau)\}$ using the second convolutional layer
    6) Downsample the signal from the second convolutional layer and obtain the output symbols which are denoted as $\{\hat{x}\}$
    7) Compute the error $L(w,h) = ||\hat{x} - x||^2$
    8) If loss does not decrease for 6 epochs
        break
        Else
        a) Backpropagate to compute gradient $\nabla_w L(w^n, h^n)$ and $\nabla_h L(w^n, h^n)$
        b) Compute
        $w^{(n+1)} \rightarrow w^n - \eta Nadam(\nabla_w L(w^n, h^n))$
        $h^{(n+1)} \rightarrow h^n - \eta Nadam(\nabla_h L(w^n, h^n))$
        c) Update trainable parameters
    End for
---

Since the original updating process of the adaptive filter, which in general employs a least mean square (LMS) algorithm, is equivalent to the optimization of the last convolutional layer of the NN, we only need to extract the weights of the first convolutional layer. This is important to ensure that the NN-based joint-transceiver equalization optimization is compatible with conventional Rx DSP. Furthermore, as Eq. (4) shows, the weights need to be flipped and then are used as the coefficients of the pre-equalization filter.

*C. Practical considerations*

The NN-based optimization requires a differentiable channel model in order to compute gradients. One approach is to utilize traditional models with necessary calibration. The coefficients of transceiver-joint equalization depend on the distribution of noise and optical filtering. The noise includes the amplified spontaneous emission (ASE) noise induced by Erbium-doped fiber amplifier (EDFA) and the fiber nonlinear noise. Traditional models suffer from both the inaccuracy in the models themselves and model input parameters, especially for WSS and EDFA. Many approaches have been proposed to calibrate the model and the model parameters. For example, a method which relies on optical spectra captured by optical channel monitoring is proposed to estimate the central frequency and the bandwidth of WSSs in [15], and a method based on NN is proposed to estimate the gain and the noise figure of EDFAs in [16]. However, in general the model accuracy still needs to be significantly improved in deployed systems to ensure the performance of the NN-based optimization.

Another approach is to reconstruct the channel using a generative adversarial network (GAN). In [17], it has been demonstrated that GAN can learn the distribution of fiber channel transfer function including attenuation, fiber nonlinearity, chromatic dispersion and ASE noise. However, fast-changing impairments such as laser phase noise and rotation of state of polarization (RSOP) are not considered and cannot be learned by GAN. In a practical system, only the output signal of the channel which has suffered from the fast-changing impairments is feasible. Therefore, the fast-changing impairments have to be compensated for before learning the channel distribution. Furthermore, since pre-equalization may change the feature of the input signal, which is also the input of GAN, the output of GAN might be different from the real output of the channel. Therefore, the generalization of GAN is challenging and requires further study.

The Tx laser frequency offset (FO) is another nonnegligible factor that influences the performance of transceiver-joint equalization. It has been reported that the Tx laser FO drifts fast over time at the MHz/s magnitude [18]. When the Tx laser FO drifts, the offset between the central frequency of the signal and the central frequency of the WSSs changes, such that the effect caused by the optical filtering varies. In this case, if we do not correct the model and retrain the coefficients of the equalization using the new model in time, the gain of the transceiver-joint equalization will degrade. To overcome this problem, it is essential to monitor the Tx laser FO and modify the model of the training link. Monitoring the Tx laser FO using the spectrum of the received signal may be a promising candidate because the Tx laser FO will be reflected by the spectrum of the received signal. Afterwards, the channel model is modified, and the coefficients of the transceiver-joint equalization are retrained using the new model. Considering the time delay to update the model and retrain the pre-equalization coefficients, residual FO deviation is inevitable. Therefore, an analysis of the effect of the Tx FO is needed, and it will be numerically studied in Section III-E.

III. SIMULATION SETUP AND RESULTS

In this section, extensive numerical results for transceiver-joint equalization of optical filtering impairments are presented to verify the proposed scheme. Two links are employed in the simulation, named as evaluation link and training link, respectively. The evaluation link is a traditional communication system used for performance evaluation. The training link is an NN equivalent to the evaluation link as described in Section II for the optimization. Fig. 3 depicts the simulation setup used for the evaluation link. At the Tx side, a 65 GBaud dual-polarization 16-QAM signal is generated with a length of $2^{15}$ symbols. A root-raised-cosine (RRC) filter is used for pulse shaping with a roll-off factor of 0.01. After that, pre-equalization is performed. The laser linewidth is 100 kHz in the Tx and Rx. In addition, -21 dB additive white Gaussian noise (AWGN) is added to simulate the noise of the Tx. Then, the signal enters the fiber channel, which consists of standard single-mode fiber (SSMF), EDFA and WSS. The fiber length of each span is 80 km. An EDFA with a noise figure of 5 dB and gain of 16 dB is located at the end of each span to completely compensate for the loss of fiber. A WSS is inserted after the EDFA in every span, and the transfer function $S(f)$ of the WSS is shown as follows:

$$S(f) = \frac{1}{2}\sigma\sqrt{2\pi}[erf(\frac{(B_0/2-f)}{\sqrt{2}\sigma}) - erf(\frac{(-B_0/2-f)}{\sqrt{2}\sigma})] \quad (11)$$

$$\sigma = \frac{B_{OTF}}{2\sqrt{2\ln 2}} \quad (12)$$

where $erf(\cdot)$ represents the Gauss error function, $B_0$ denotes the bandwidth of the WSS, $f$ denotes the frequency, and $B_{OTF}$ denotes the steepness of the edge of the transfer function. In this paper, $B_{OTF}$ and $B_0$ are 73 GHz and 8.8 GHz, respectively. Similarly, -21 dB AWGN is added to simulate the noise of the Rx. In this paper, we consider three link scenarios: 1) only fiber loss is considered; 2) fiber nonlinearity and CD are introduced; 3) polarization effects are further included. At the Rx, we implement a least-mean-square (LMS) based adaptive filter combined with a phase-locked loop (PLL). Afterwards, the SNR is calculated. Note that CD compensation is only conducted in the second scenario. The simulation parameters are summarized in Table III.

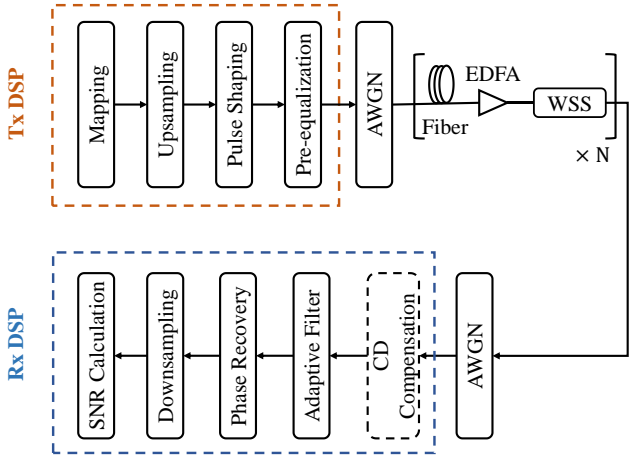

Fig. 3. The simulation setup of the evaluation link.

TABLE III
PARAMETERS OF SIMULATION

| Parameter | Value |
|---|---|
| Baud rate | 65 GBaud |
| Symbol length | $2^{15}$ |
| Sample per symbol | 2 |
| RRC roll off factor | 0.01 |
| Laser linewidth | 100 kHz |
| Tx AWGN | -21 dB |
| Launch power | 0 dBm |
| EDFA gain | 16 dB |
| EDFA noise figure | 5 dB |
| WSS bandwidth (-6 dB) | 73 GHz |
| WSS steepness of edge | 8.8 GHz |
| Fiber loss | 0.2 dB/km |
| Fiber dispersion | 16.7 ps/(nm·km) |
| Fiber span length | 80 km |
| Kerr coefficient | 0.0013 $w^{-1} \cdot m^{-1}$ |
| SOP speed | 1e5 rad/s |
| PDL | 3 dB |
| PMD | 40 ps |
| Rx AWGN | -21 dB |

As for the training link, the Tx pre-equalization filter and the Rx adaptive filter of the evaluation link are replaced with two convolutional layers, respectively. Phase noise and phase recovery algorithm are not included in the training link since it does not affect the optimal transceiver-joint equalization coefficients. The rest of the configuration is the same as the evaluation link. The setup of the training link is built in Pytorch to perform the optimization of the parameters while the performances are evaluated using the simulation of the evaluation link in MATLAB. In order to use complex gradient descent, we set the real and imaginary parts as the two dimensions of the signal during training. The parameters of the Tx and Rx convolutional layers are the same and they are summarized in Table IV. The kernel size is 101, which equals to the length of the pre-equalization filter and adaptive filter. The number of padding is 50 to guarantee that the input and output of convolutional layer have the same length. The strides of all the convolutional layers are 1.

TABLE IV
PARAMETERS OF THE CONVOLUTIONAL LAYERS

| Parameter | Value |
|---|---|
| Input channels | 2 |
| Output channels | 1 |
| Kernel size | 101 |
| Padding | 50 |
| Groups | 1 |
| Bias | False |

The training procedure is conducted as described in Algorithm 1. The NN is trained using the Adam optimizer with a learning rate of 0.001. The optimization is conducted for 150 iterations and early stop is used. As the noise of EDFA and transceiver are randomly generated at each training iteration, no batch is identical during the training. This prevents overfitting problems that may arise when using a training dataset of a fixed and limited size that is reused multiple times during the training process [12].

After training, we evaluate the performance of the transceiver-joint equalization using the evaluation link. We extract the weights of the first convolutional layer and denote it as $w$. As mentioned above, $w$ needs to be flipped and then utilized as the pre-equalization filter coefficients of the evaluation link. Note that the implementation of the adaptive filter in the Rx DSP is not changed.

### A. The results when only fiber loss is considered

In this section, only the fiber loss is considered while fiber nonlinearity and CD are not introduced. We simulate two scenarios. The first scenario is that all the WSSs have no FO and the second scenario is that all the WSSs have a 2 GHz FO.

To train and test the NN, a sufficient number of datasets are first generated. The dataset consists of 7680 sets and each set contains 512 symbols. The dataset is divided into the training set (6400 sets) and the testing set (1280 sets). The loss curves on the training and testing sets of the system are shown in Fig. 4. In order to show the optimization process more clearly, no initialization of the NN is conducted in this step. It can be observed that the system has converged, and no overfitting occurs during the training process. The insets of Fig. 4 depict the spectrums of the weights of the first convolutional layer in the initial stage or the finished stage of the training.

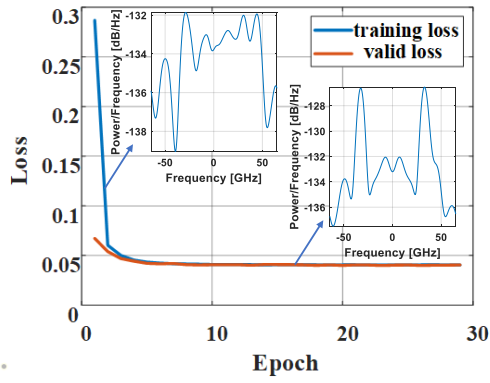

Fig. 4. The learning curve of the NN.

Fig. 5 illustrates the spectrum of the signal before and after the pre-equalization when the number of WSSs is 16 and all the WSSs have no FO. We can observe that the energy of the signal has been reallocated in the frequency domain, as a result of which, the high frequency components of the signal which suffer strong filtering impairments are emphasized.

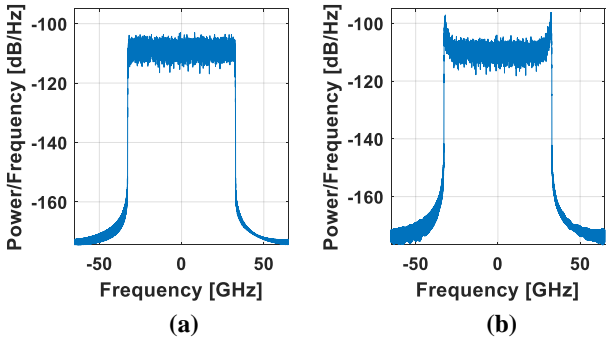

Fig. 5. The spectrum of the signal (a) before pre-equalization and (b) after pre-equalization when the number and FO of WSS are 16 and 0, respectively.

Fig. 6 indicates the SNR versus the number of WSSs in the link. The number of WSSs ranges from 10 to 16, and the coefficients of the transceiver-joint equalization are retrained when the link varies. The benchmark is the performance with receiver equalization only. As shown in Fig. 6, the gain of the transceiver-joint equalization increases as the number of WSSs increases. When the number of WSSs is 16, the SNR is improved by 0.76 dB.

Fig. 7 depicts the SNR versus the number of WSSs in the link when the FOs of all the WSSs are 2 GHz. Unlike the scenario without FO, the gain of the transceiver-joint equalization does not increase as the number of WSSs increases. When the number of the WSSs is 10, the SNR gain is 0.94 dB, but when the number of WSSs is 16, the SNR gain is only 0.64 dB. The main reason is that for a larger number of WSSs, signal power becomes lower after pre-equalization in the transmitter due to the low-pass filtering caused by the transmitter frequency response. In addition, as the pre-equalized signal carries more power in the high frequency components, more signal power is carved out after propagating through WSSs, thereby degrading the performance of transceiver-joint equalization.

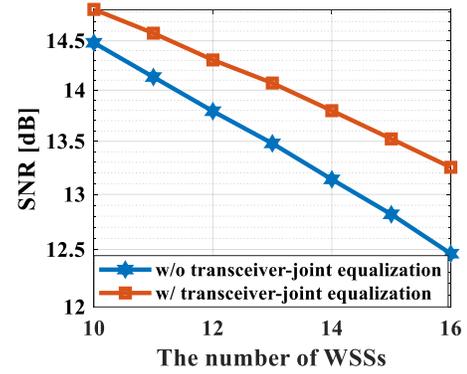

Fig. 6. SNR versus the number of WSSs when the FOs of WSSs are 0 GHz.

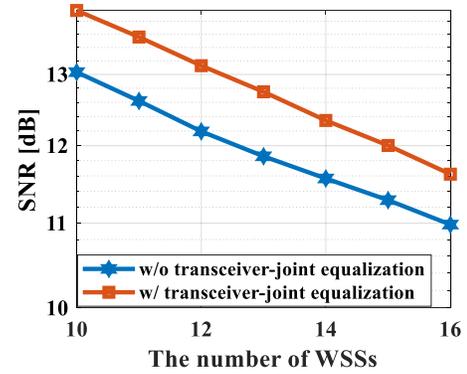

Fig. 7. SNR versus the number of WSSs when the FOs of all the WSSs are 2 GHz.

### B. The results when fiber nonlinearity and CD are introduced

In this section, the scenario considering fiber nonlinearity and CD is simulated. The fiber is simulated using SSFM. To guarantee the accuracy of the SSFM, a constant step size of 10 m is used. It is worth mentioning that the FOs of all the WSSs are 0 GHz.

At the Rx, CD compensation is first performed before the adaptive filter. Since additional nonlinear noise induced by fiber nonlinearity changes the existed distribution of filtering impairments and noise, the coefficients of the optimal transceiver-joint equalization will change. Thus, the fiber nonlinearity model needs to be included in the training link. Furthermore, since the additional nonlinear noise changes the ratio between noise and filtering impairments, the gain of the transceiver-joint equalization is different from that of the situation without fiber nonlinearity as shown in Fig. 8. For example, when the number of WSSs is 16, the SNR gain is 0.83 dB, larger than 0.76 dB in the previous situation.

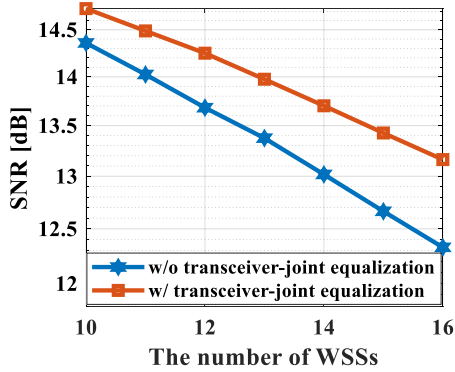

Fig. 8. SNR versus the number of WSSs when fiber nonlinearity and CD exist.

### C. The results when polarization effects exist

In this section, we test the scenario when polarization effects exist. In this case, the fiber nonlinearity and CD are neglected. Note that polarization effects are only added in the evaluation link, and they are not considered in the training link. This is because polarization impairments are often compensated in the Rx and they do not affect the coefficients of the pre-equalization. And in our scheme, we only utilize the coefficients of the pre-equalization extracted from the NN. The polarization effects are composed of polarization dependent loss (PDL), RSOP and polarization mode dispersion (PMD). The polarization model can be denoted as follows:

$$J(\omega) = \begin{bmatrix} \sqrt{1+\gamma} & 0 \\ 0 & \sqrt{1-\gamma} \end{bmatrix}$$
$$\cdot \begin{bmatrix} \cos\alpha e^{j\xi} & -\sin\alpha e^{j\eta} \\ \sin\alpha e^{-j\eta} & \cos\alpha e^{-j\xi} \end{bmatrix} \cdot \begin{bmatrix} e^{j\omega\tau/2} & 0 \\ 0 & e^{-j\omega\tau/2} \end{bmatrix} \quad (13)$$

where the first, second and third matrix denotes the PDL, RSOP and PMD matrix, respectively. $\gamma$ is the magnitude of the PDL vector which is defined as the ratio between the maximum and minimum attenuation. PDL in dB is denoted as follows:

$$PDL(dB) = 10\log_{10}(\frac{1+\gamma}{1-\gamma}) \quad (14)$$

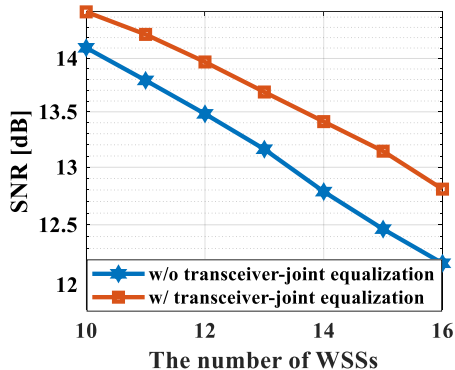

Fig. 9. SNR versus the number of WSSs when polarization impairments exist.

The RSOP matrix including 3 parameters $\alpha, \xi, \eta$ can fully describe the rotations in the fiber channel. $\tau$ in the PMD matrix denotes the differential group delay (DGD). Since PDL will make the performance of the system degrade, as shown in Fig. 9, the SNR of the system declines when the additional polarization impairments are introduced. Nonetheless, when the polarization impairments exist, the gain of the optimal joint-transceiver equalization is almost the same as that without polarization effects, indicating that the scheme we propose is robust against polarization effects.

### D. The results of two different distributions of filtering impairments and noise

Since the gain of the transceiver-joint equalization is relevant to the distribution of filtering impairments and noise, two different scenarios are considered in the section. In the first scenario, most of the noise is generated after optical filtering, whereas in the second scenario, most of the noise is filtered by the cascaded WSSs. The schematics of the two scenarios are shown in Fig. 10. The number of WSSs are 10 and the number of spans are 15 in both scenarios and only the distribution of the WSSs and spans are different.

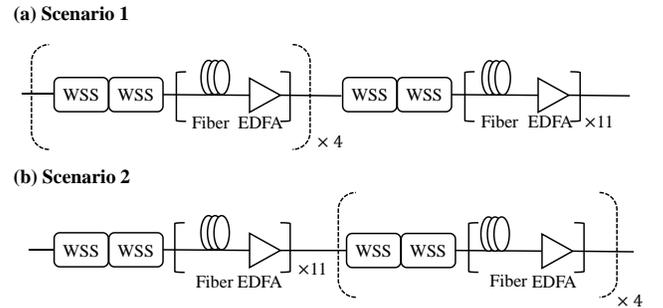

Fig. 10. The link setup of the two scenarios.

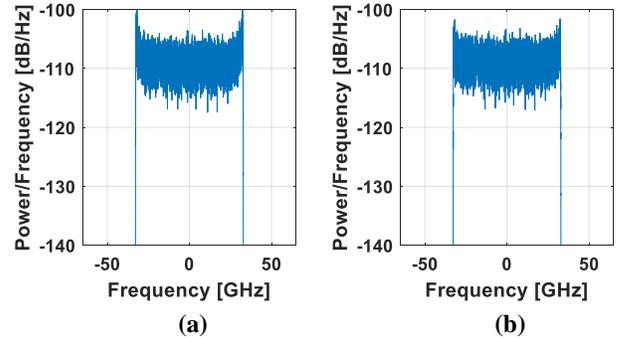

Fig. 11. The spectrums of the signal after pre-equalization when most of the noise is (a) after and (b) before optical filtering.

The spectrums of the signal after the pre-equalization in the two scenarios are shown in Fig. 11. When the noise is added after the filtering impairments, the high frequency components of noise are enhanced when equalization is only conducted in Rx. Thus, stronger equalization needs to be conducted in the Tx. On the contrary, when the noise is added before the filtering impairments, the noise will not be enhanced by the Rx equalization, and the pre-equalization will cause the excess power loss of the signal after filtering. Therefore, stronger equalization needs to be conducted in the Rx. In Scenario 1, more noise is added after filtering impairments than Scenario 2, so stronger equalization is conducted in the Tx and more power is allocated to the high frequency components after pre-equalization as shown in Fig. 11. Furthermore, in Scenario 1, the gap between state without transceiver-joint equalization and the optimal transceiver-joint equalization is large than that of Scenario 2. Thus, the scenario 1 achieves more gain than the scenario 2, as shown in Table V.

TABLE V
THE SIMULATION RESULTS OF DIFFERENT LINK DISTRIBUTIONS

| Scenario | W/o transceiver-joint equalization | W/ transceiver-joint equalization | Gain [dB] |
| --- | --- | --- | --- |
| Scenario 1 | 13.34 | 13.75 | 0.41 |
| Scenario 2 | 13.82 | 14.11 | 0.29 |

*E. The evaluation of the Tx FO effects*

As explained in Section II-C, the Tx FO influences the performance of the transceiver-joint equalization. In this section, we simulate three different equalization schemes. In the first scenario, the equalization is only conducted in the Rx. In the second and third scenarios, transceiver-joint equalization is conducted, with an accurate Tx FO value and a fixed FO of 0 GHz, respectively. The number of the WSSs is 10 and all the WSSs have no FO. The simulation result is shown in Fig. 12. It is observed that when the center frequency of Tx drifts, the gain of the pre-equalization decreases. When the Tx FO is 2.5 GHz, the gain of the pre-equalization is reduced by 0.61 dB. Therefore, in order to achieve the best performance, it is essential to monitor the Tx FO drifting and modify the model to retrain the coefficients of the transceiver-joint equalization. As shown in Fig. 12, when the Tx FO is 1 GHz, the SNR gain degrades by only 0.08 dB. This means that the tolerance of the Tx FO deviation can reach 1 GHz, which makes the implementation of the transceiver-joint equalization more feasible.

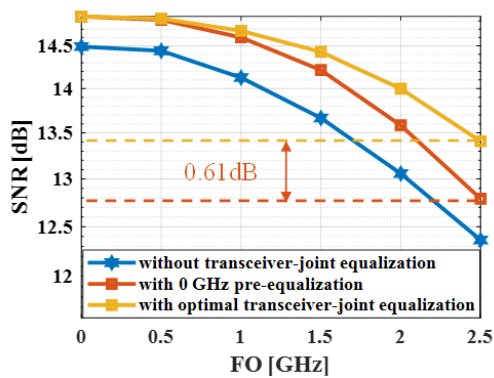

Fig. 12. SNR versus the Tx FO drifting with different pre-equalization schemes.

## IV. CONCLUSION

In this paper, we propose an optimal transceiver-joint equalization scheme for compensating for the WSS-induced filtering impairments by leveraging the optimization ability of AI. Instead of introducing a black box, we exploit the connection between a communication system and an NN. By mapping a communication system to an NN and optimizing the mapped NN, the optimal transceiver-joint equalization of the communication system can be obtained. Since the equalization coefficients of Rx can be obtained adaptively in a practical system, only the coefficients of the pre-equalization need to be extracted and the DSP structure of the communication system is not changed. Through extensive simulations, we demonstrate the good performance of the proposed scheme when various fiber impairments including nonlinearity, CD and polarization effects are considered. In addition, we analyze the performance of different distributions of noise and optical filtering impairments. Finally, the impact of Tx FO on the performance of the transceiver-joint equalization are evaluated.